# Entry effects of droplet in a micro confinement: implications for deformation-based CTC microfiltration

Zhifeng Zhang[1], Xiaolin Chen[1], Jie Xu[2,a]

1. Department of Mechanical Engineering, Washington State University, Vancouver, WA, 98686, USA

2. Department of Mechanical and Industrial Engineering, University of Illinois at Chicago, Chicago, IL 60607, USA

Deformation based circulating tumor cell (CTC) microchips are a representative diagnostic device for early cancer detection. This type of device usually involves a process of CTC trapping in a confined microgeometry. Further understanding of the CTC flow regime, as well as the threshold passing-through pressure is key to the design of deformation based CTC filtration devices. In the present numerical study, we investigate the transitional deformation and pressure signature from surface tension dominated flow to viscous shear stress dominated flow using a droplet model. Regarding whether CTC fully blocks the channel inlet, we observe two flow regimes: CTC squeezing and shearing regime. By studying the relation of CTC deformation at the exact critical pressure point for increasing inlet velocity, three different types of cell deformation are observed: 1) hemispherical front, 2) parabolic front, and 3) elongated CTC co-flowing with carrier media. Focusing on the circular channel, we observe a first increasing and then decreasing critical pressure change with increasing flow rate. By pressure analysis, the concept of optimum velocity is proposed to explain the behavior of CTC filtration and design optimization of CTC filter. Similar behavior is also observed in channels with symmetrical cross sessions like square and triangular but not in rectangular channels which only results in decreasing critical pressure.

## I. INTRODUCTION

Cancer is a major public health problem all over the world[1]. Decades of efforts have been engaged in controlling cancer death rates with only marginal success. According to National Cancer Institute (NCI) database, cancers can be categorized into local stage, regional stage and distant stage[2]. Early detection at local stage has shown a high five-year survival rate for many kinds of cancers such as cervical cancer [3], prostate cancer[4], pancreatic cancer[5] and *etc*. Detecting cancers early when treatment is most successful offers substantial benefits to patients.

There is growing evidence that circulating tumor cells (CTCs) present in the bloodstream of cancer patients are responsible for initiating cancer metastasis and can be used as a potential biomarker for early cancer detection[6-8]. CTCs are tumor cells shed from primary tumors and can travel through circulation to distant metastatic sites or even self-seed into their tumors of origin [9].

---

a Corresponding author email: jiexu@uic.edu

The detection of CTCs has important prognostic and therapeutic implications as they are found in the peripheral blood of early-stage cancer patients prior to of the onset of clinical symptoms[8]. A variety of technologies have been developed to separate CTCs from normal blood cells, based on the differences in cellular biochemical properties such as surface antigen expression or physical properties, such as size or deformability[10] as well as differing acoustic[11, 12], magnetic[13], optic[14] or dielectrophoretic characteristics [11, 15-18] et al. Among the various extant methods for CTC detection and enumeration, deformation-based CTC filtration offers the advantages of structural simplicity[19, 20], stable performance and low cost[21]. For example, deformation-based CTC microfilters, such as orifice[22], weirs[20], ratchet[23], pillars[24], or membranes[21]-based devices, can batch-process samples of whole blood without elaborate chemical labelling[25], and provides a compelling label-free[26] strategy for blood purification and cancer cell capturing.

Most existing studies on deformation-based CTC microfilters are experimental work focusing on device design and manufacturing[15, 27-29]. In order to achieve high capture efficiency, high isolation purity and high system throughput[27, 30], CTC microfilter designs need to be optimized for performance. Numerical modelling can play a crucial role in increasing our ability to understand the complex cell/flow/channel interactions in the cell entrained pressure-driven flow passing through microfluidic channels, and may deliver important insights into optimum design of future lab-on-a-chip devices.

Previous numerical studies have employed a liquid-droplet model for cells and reported findings on entry channel pressure influences for semi-static device design[16], with relevant pressure and deformation correlation[31, 32], as well as the effect of 3D channel geometry on cell deformation [16]. The pressure signature[33] and possible deformation evolution under the influence of both surface tension of the cancer cell and viscous shear stresses acting on the cell at different device operating conditions remain elusive. Transitional operating conditions between surface tension-dominant and viscous shear stress-dominant flow behaviors are also characterised. Considerations important to the device design include achieving maximum filtration capability, high throughput, and high efficiency. Understanding of the interplay between the effects of interfacial, viscous and inertial forces, as well as the balance between the viscous shear stressviscous shear stress and surface tension [34] is essential to develop a microfilter capable of efficient, specific, and reliable capture of CTCs.

In this paper, we study the deformation of CTC, the pressure signature of CTC passing through the microfilter, and cross-section influence over a wide range of flow rates. The pressure signature of a typical CTC deformation process is investigated for four types of 3-D geometries with uniform lengthwise cross-sectional areas: circular, square, triangular, and rectangular. The numerical study are based on Continuum Surface Force (CSF) and Volume of Fluid (VOF) methods[35]. The dynamic variance of CTC passing-through pressure is studied and compared for different channel geometries. These data and analyses provides useful insight not only for high efficiency CTC microfilter development, but also aids in understanding capillary blockage [36], cancer metastasis[37], and drug delivery processes[38].

## II. PROBLEM DESCRIPTION

CTCs are generally stiffer than normal blood cells[39] although malignant CTCs with increased metastatic potential tend to have increased deformability[40, 41]. Isolation of CTCs using deformation is achieved as the stiffer CTCs are quickly trapped in narrow channels when a fluid sample is injected into a microfluidic system. In the present study, we consider filtering channels with four distinct cross sectional shapes, namely circular, square, rectangular, and triangular. Effective CTC capturing has been reported to have a critical diameter in the range of 5-12 μm for circular microfilters[20, 42]. A circular channel with a diameter of 5 μm is adopted as the baseline case in this study for achieving high filtering performance. The key dimensions of non-circular channels are determined based on arriving at the same pressure drop in the viscous channel flows as the circular baseline calculation. Fig. 1 illustrates schematically the microfiltration setup. Carrier flow and CTC along the microchannel proceeds from the left to the right. The device is working on the characteristic dimensions on the order of 10 μm.

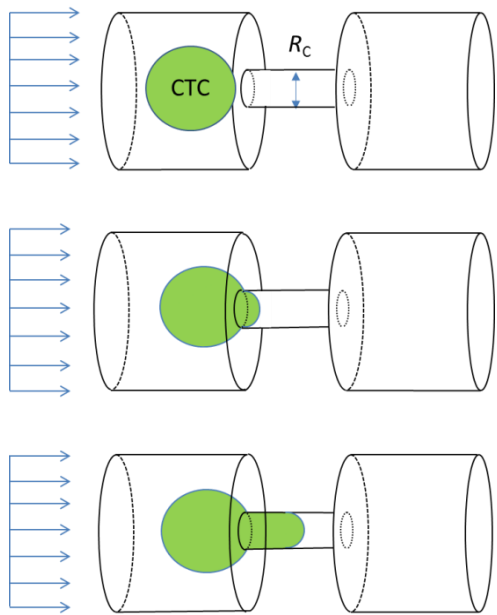

Fig. 1 A cell passing through a filtering channel.

We model a cell, which is an aqueous core surrounded by a thin lipid bilayer membrane, as a liquid droplet. In the droplet model, the cell's ability to resist deformation is described by the surface energy (interfacial tension) of the droplet. CTCs derived from different cancer types are investigated. Table 1 gives the cancer cell types and their corresponding interfacial tension coefficients extracted from single cell deformability experiments reported in the literature. Note that a more deformable cell possesses a lower interfacial tension coefficient. As expected, surface tension of white blood cells (WBCs) is substantially lower than those of the stiffer cancer cells listed in Table 1. The droplet model adopted here neglects the cell internal contents such as nucleus and microfilaments and does not account for heterogeneity of the cell, nor does it allow the description of active responses exhibited by the living cell, such as rigidity sensing or cell pulsation. However, it readily allows insights into the filtration effects of different cancer cell types under various operating conditions for cell capturing. Indeed, previous studies also show that droplet model is suitable for simulating cells especially when the deformation is large [43].

Table I Cell types and their interfacial tension coefficients

| Cell type | Cell line | Interfacial tension coefficient (mN/m) | Ref |
|---|---|---|---|
| Neutrophils(WBC) | - | 0.027 | [44] |
| Macrophage | - | 0.14 | [45] |
| Myeloid cell | HL60 | 0.15 | [45] |
| Brain cancer | U-87 Mg | 7 | [46] |
|  | LN-229 | 10 |  |
|  | U-118MG | 17 |  |
| Esthesioneuroblastoma | ENB | 20 | [47] |
| Salmonella Enteritidis | - | 40 | [48] |
| Cervical cancer | - | 50 | [41] |

The size distribution of CTCs (12-25 μm in diameter has shown to overlap with that of normal WBCs (10-18 μm in diameter [44]). In our model, CTCs and WBCs, are both considered to be 16 μm in diameter, to rule out effects resulting from the size differences in cells. The erythrocyte (7-8 μm in diameter [44]) are not considered in the study as they are highly deformable and can pass freely through any filtering channel within the design range.

## III. MODELING OF TWO-PHASE FLOW WITH SURFACE TENSION

The problem under consideration consists of a dispersed phase (the cell) entrained in a continuous phase (the ambient flow) passing through a microfluidic contraction [49, 50]. The interface that separates the two phases can be solved using volume of fluid (VOF) [16, 51], level-set [43], front-tracking [52] or phase-field method. Among these methods, the VOF method, which has superior mass conservation property, is adopted in our model to track the interface between the cell and its surrounding fluid. Volume of Fluid (VOF) method uses simple and economical way to track free surface[53]. It defines and stores only a function $\alpha$ whose value is 1 at any point occupied by CTC, and 0 otherwise. The value between 0 and 1 indicates the membrane of CTC.

Transient simulation of a cell passing through a filtering channel is performed using the commercial software ANSYS Fluent with an explicit time stepping scheme. Inflation layers are employed to properly resolve the flow within the boundary layer. The 3-D mesh is inflated along the walls of the non-circular channels. The circular channel is modeled using a 2-D axisymmetric model, with inflated mesh distributed along the symmetry line as well as the channel walls. The total number of elements used for the 3-D simulation is approximate 200,000, with nearly 5000 elements patched to the cell. Both CTCs and white blood cells are assumed to be incompressible.

The interface is tracked by solving the transport equation for the dispersed phase volume fraction α. Assuming no mass transfer between phases, the continuity equation can be written as:

$$\frac{\partial \alpha}{\partial t} + \nabla \cdot (\alpha \vec{V}) = 0 \tag{1}$$

Besides the interface tracking, simulation needs to take into account surface tension effects. Our surface tension model is based on the CSF method. With this model, the addition of surface tension to the VOF calculation results in a source term in the conventional incompressible Navier-Stokes momentum equation[35]:

$$\frac{\partial}{\partial t}(\rho \vec{V}) + \nabla \cdot (\rho \vec{V} \vec{V}) = -\nabla p + \nabla \cdot [\mu(\nabla \vec{V} + \nabla \vec{V}^T)] + \vec{F} \tag{2}$$

where $\vec{V}$ is the velocity vector. $\rho$ and $\mu$ are volume-fraction-averaged density and viscosity, respectively. $\vec{F}$ is the source term representing the surface tension force localized at the interface.

The surface tension force can be transformed into a volume force in the region near the interface, and expressed as [35]:

$$\vec{F} = \sigma \frac{2\rho \kappa \nabla \alpha}{(\rho_1 + \rho_2)} \tag{3}$$

where $\sigma$ is the interfacial tension coefficient between the dispersed phase and the continuous phase, $\kappa$ is the mean curvature of the interface in the control volume. $\rho_1$ and $\rho_2$ are densities of the continuous phase and the dispersed phase, respectively. The density is interpolated from $\rho = \alpha \rho_1 + (1 - \alpha) \rho_2$.

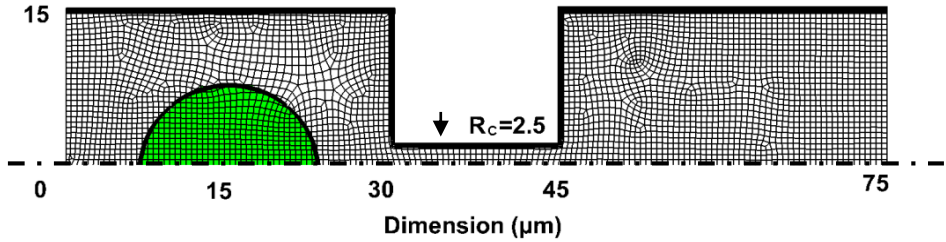

Fig. 2 Mesh and dimensions for circular channel. The CTC locates at 15 μm away from the filter inlet. Area patched with the volume fraction of 1 indicates the CTC.

For the incompressible two-phase flow problem, a constant volume flow rate is used as the inlet boundary condition. No-slip boundary conditions are prescribed on the channel walls. The walls are assumed to be non-wetting with respect to the dispersed phase with a 180 degree contact angle.

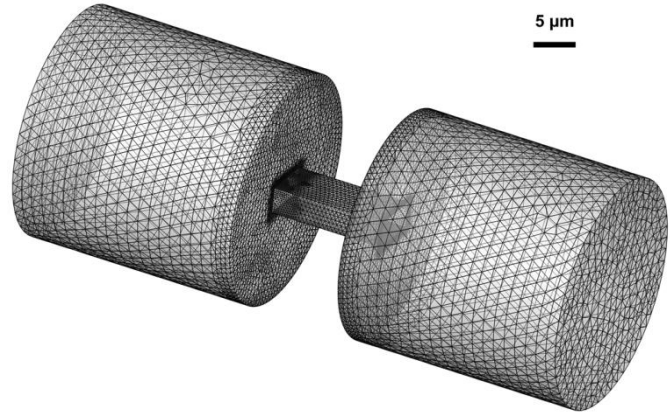

Fig. 3 Mesh and dimensions for the non-circular microfilter (rectangular)

## IV. RESULTS AND DISCUSSION

In the literature, the problem of modelling a fluid body passing through a channel confinement has mainly focused on either the influence of surface tensionor viscous shear stress on the cell or droplet[54]. Briefly, droplet motion at low velocity can be described by surface-tension-dominated flow while high-speed motion can be described as shear-stress-dominated flow. The dimensionless numbers that can be used to describe the relations of important forces include *Ca* and *We* number as defined in Eq. (4).

For low flow rate cases, the threshold pressure of CTC passing through a microchannel is determined by the Young-Laplace equation assuming quasi-static conditions. Examples of slow motion application of the Young-Laplace equation include micropipette aspiration [55] and ratchet CTC chips [19, 23]. For high flow rate cases, research has been focused on the friction- and viscosity-induced non-linear droplet transport pressure.

### A. CTC DEFORMATION BEHAVIOR AT DIFFERENT OPERATIONAL CONDITIONS

Circulating tumor cells can be categorized into three types by their stiffness rigid, medium and compliant According to the list of interfacial tension coefficients in Table 1. Together with extremely soft WBC, the flow regimes of four types of cells are illustrated in Fig 4. Three characteristic stages are selected to represent different flow regimes.

A flow regime chart based on dimensionless numbers is often used in multiphase flow study [56]. Here we choose the x-axis of the flow regime as the *We* number, and the Y-axis is selected as the *Ca* number, which makes the diagonal lines tilted 45 degree the *Re* number. These dimensionless numbers are defined below

$$We = \frac{\rho V^2 d}{\sigma} \qquad Ca = \frac{\mu V}{\sigma} \qquad Re = \frac{\rho V d}{\mu} \qquad (4)$$

Here, *d* is the diameter of confinement channel, *ρ* is the density of cell, *V* is the velocity of carrier flow, *σ* is the interfacial tension coefficient of CTC, and *μ* is the viscosity of carrier flow.

A series of CTC deformation simulations was conducted as listed in Fig. 4. The W, S, M, H represents white blood cells, CTCs (compliant), CTCs (medium) and CTCs (rigid), respectively. Three stages were plotted and illustrated. For purpose of pressure analysis we are most interested in the first stage, where the maximum threshold pressure is reached. As can be seen for CTCs (rigid), we illustrated the cell deformation at four different flow rates. (H1) deformation is a representative of hard cell passing through the microfilter at an extremely low velocity. Since the viscous shear stress influence is negligible, the deformation of the cell front at critical pressure follows a spherical shape and the cell is squeezed into the confinement channel. (H4) deformation shows an elongated cell front, and the cell is detached from the wall of the microchannel, resulting in co-flow of the carrier fluid inside the filtering channel. In this case, the cell is sheared into the confinement channel. Similar phenomena can be observed for other classes of cell stiffness. Deformation at (W1), (S1) and (M1) represents cells being squeezed through the channel without co-flow of carrier fluid. Design at this stage should focus on CTC experiencing "friction" [40] with the wall of channel [57]. With increasing velocity, for deformation at (W2), (S2) and (M2), the carrier fluid begins to co-flow with CTC resulting in shear dominated flow. Fig. 5 further illustrates the detailed flow streamlines at the stage when threshold pressure is reached at three different flow rates for hard CTC.

If the velocity is small enough, the CTC deformation follows spherical deformation when passing through the channel as in Fig. 5 (a). The cell is passing with wall friction [51] at a low *Re* number. This is the so called "squeezing regime", and the process of CTC passing is similar to the micropipette aspiration that has been described previously [55, 58]. With increasing *Re*, shear force begins to dominate, and the front end of CTC will follow a parabolic shape as in Fig.

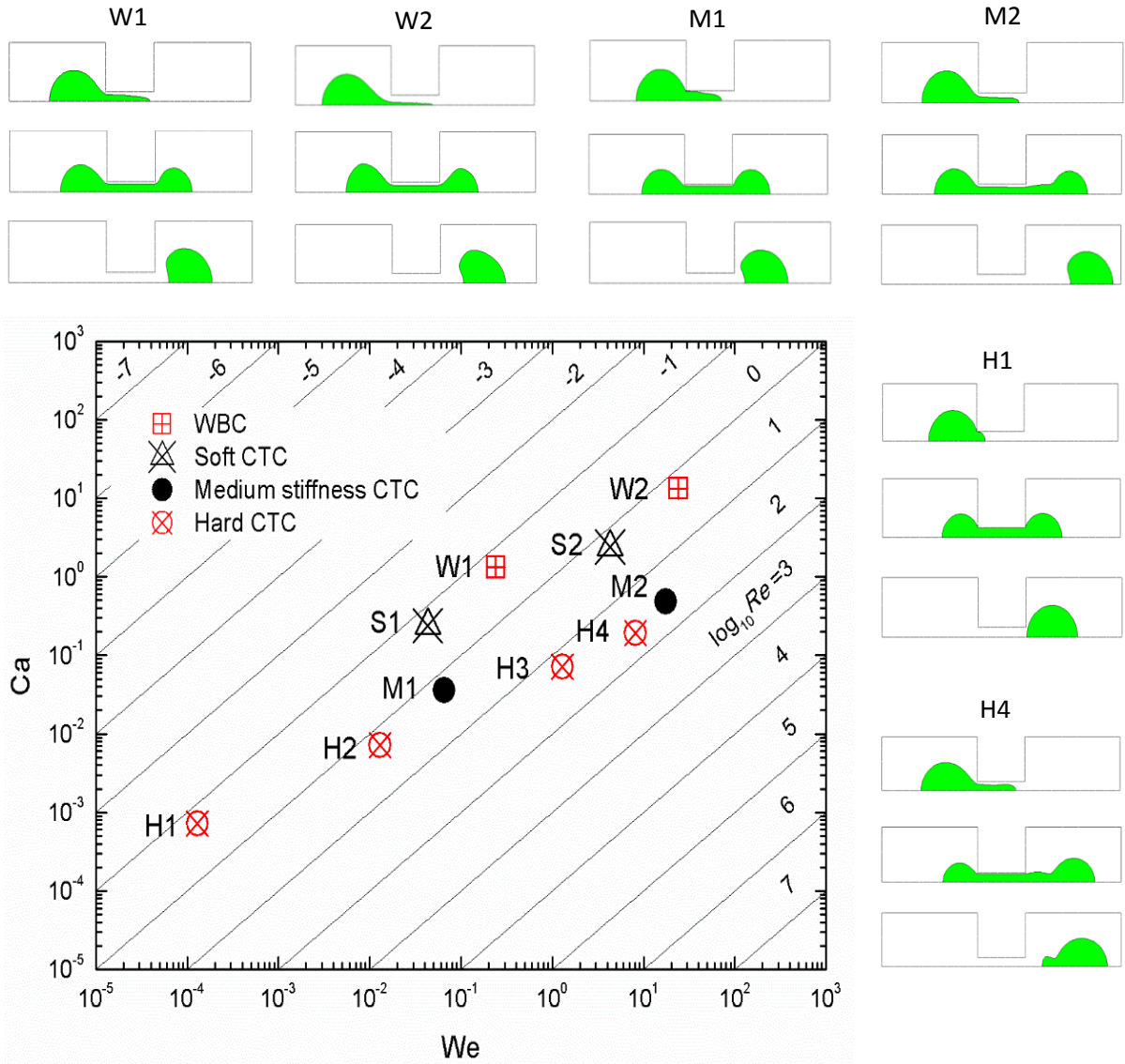

Fig. 4 Phase chart of CTC deformation behavior
(H1) Hard CTC at *Re*=0.18, *We*=1.3×10$^{-4}$, flow rate 0.7nL/s; (H2) Hard CTC at *Re*=1.8, *We*=1.3×10$^{-2}$ flow rate 7nL/s; (H3) Hard CTC at *Re*=18, *We*=1.3, flow rate 70 nL/s; (H4) Hard CTC at *Re*=36, *We*=5.2, flow rate 140nL/s; (M1) Medium stiffness CTC at *Re*=0.18, *We*=6.48×10$^{-4}$ flow rate 0.7nL/s; (M2) Medium stiffness CTC at *Re*=3.6, *We*=6.48×10$^{-2}$, flow rate 14nL/s; (S1) Soft CTC at *Re*=0.18, *We*=4.3×10$^{-2}$, flow rate 0.7nL/s; (S2) Soft CTC at *Re*=1.8, *We*=4.3, flow rate 7nL/s; (W1) WBC at *Re*=0.18, *We*=2.4×10$^{-1}$ (W2) WBC at *Re*=1.8, *We*=24, flow rate 7nL/s 8 (b). With even higher *Re*, a complete cell-wall detachment will happen, a situation that helps high speed cell passing, and the CTC front end becomes extremely thin and elongated.

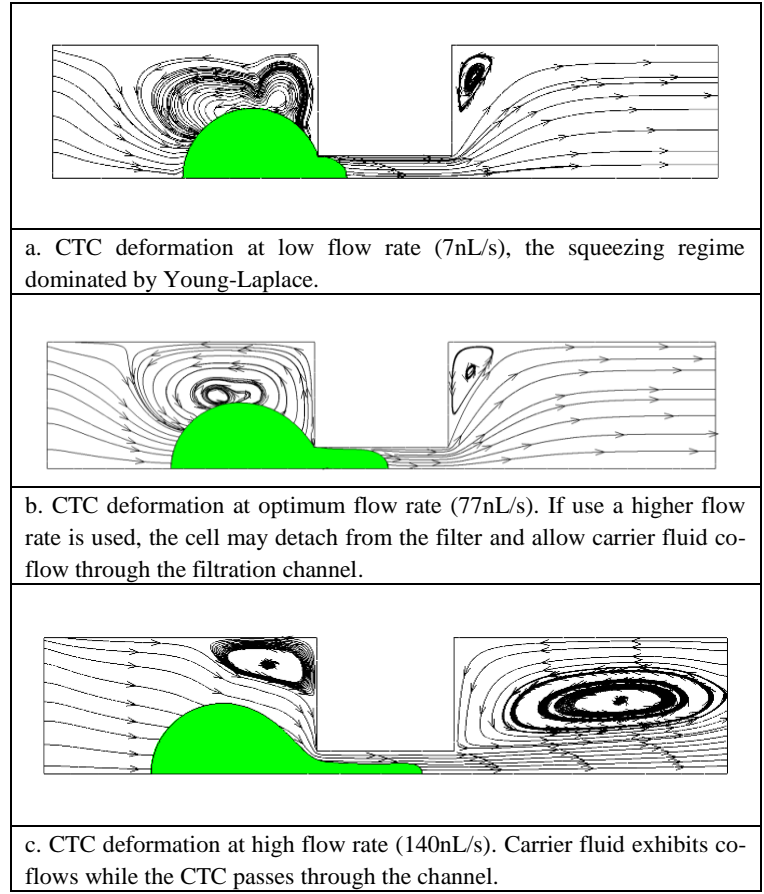

a. CTC deformation at low flow rate (7nL/s), the squeezing regime dominated by Young-Laplace.

b. CTC deformation at optimum flow rate (77nL/s). If use a higher flow rate is used, the cell may detach from the filter and allow carrier fluid co-flow through the filtration channel.

c. CTC deformation at high flow rate (140nL/s). Carrier fluid exhibits co-flows while the CTC passes through the channel.

Fig. 5 Streamline and CTC deformation behavior for flow rate of 7nL/s, 77nL/s and 140 nL/s for cervical cancer, respectively.

## B. MAXIMUM PASSING PRESSURE AT DIFFERENT OPERATIONAL CONDITIONS

Deformation-based CTC detection uses pressure-driven flow for cell capturing. For static pressure analysis [16], the total inlet pressure, $P_t$, during a cell passing through a channel is mainly used to overcome two types of resistance, 1) viscous resistance $P_{vis}$ and 2) resistance caused by cell deformation $P_F$ (due to surface tension):

$$P_t = P_{vis} + P_F \tag{5}$$

The viscous pressure involves major and minor losses[59]:

$$P_{vis} = \frac{8\eta L}{R_C^2}V + (K_C + K_E)\frac{\rho V^2}{2} \tag{6}$$

Here, the first term on the right is the viscous pressure drop along the channel, *η* is the viscosity coefficient. *L* is the length of microfilter channel. Besides $K_C$ and $K_E$ are the constriction and expansion coefficient respectively.. *V* is the average flow velocity in the filtering channel.

The other term $P_F$ is the cell deformation induced pressure drop or the filtration pressure of CTC, which is the difference between total pressure and the background pressure. What we are interested in is the critical pressure- the maximum filtration pressure in the process of CTC passing is the. At low velocity or semi-static cases, the critical pressure can be considered as the cell surface tension term predicted by Young-Laplace equation. However, with increasing of velocity, the difference between critical pressure and surface tension needs to be determined.

Further simulation observation collects the CTC deformation and filtration critical pressure together at a wide range of flow velocity in Fig. 6. As seen in Fig. 6, with increasing of flow rate, there is a maximum critical pressure for both cells with interfacial tension coefficients of 10 mN/m and 50 mN/m. Similar behavior is observed for cells with different interfacial tension coefficients.

Illustrated in Fig.6, we can understand that the critical pressure is primarily dominated by the CTC surface tension, which implies that critical pressure increase is mainly caused by the increase in leading-edge curvature. (e.g. from spherical to parabolic-like shape). With increasing flow speed, the cell will start to detach from the channel wall, and the critical pressure reaches a maximum value – we define this flow velocity as optimum velocity ($V_{cr}$), which is dependent on cell type and surface tension.

Fig. 6 Critical pressure with the influence of channel flow rate. The deformations are, a) critical pressure at low flow rate, b) critical pressure at optimum velocity, and 3) critical pressure at high flow rate with CTC co-flow with the carrier fluid. Insets (a) (b) and (c) show a schematic

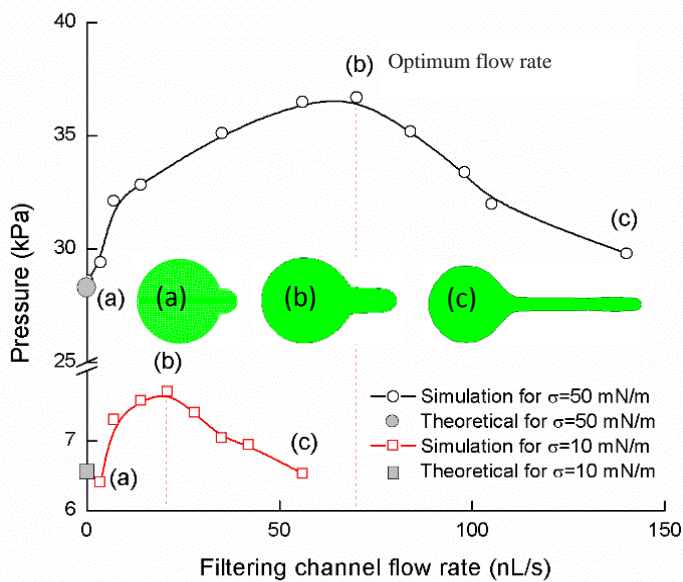

plot of CTC deformation on each characteristic stage.

At optimum velocity, the dynamic pressure in the carrier fluid is just strong enough to overcome the surface tension force generated by the CTC front. After the velocity is higher than the optimum velocity, the carrier fluid pressure will overcome the surface tension of CTC. Consequently, the carrier fluid will co-flow with the CTC. With a further increase in flow rate, the gap between the CTC and the microfilter wall becomes larger, CTC passthrough resistance, as well as the filtering capabilities of the device decreases accordingly.

## C. CELL SPECIFIC OPTIMUM OPERATIONAL CONDITIONS

Motivated by these observations, we wish to find the maximum filtering ability of a microfilter for certain types of CTC for earlier and more specific CTC detection. According to Fig.6, very soft CTCs with a low interfacial tension coefficient follow the carrier flow well when entering the microfilter channel, and accordingly, soft-CTC radius of curvature at the filter is difficult to estimate. CTCs (rigid) maintain a spherical front well under pressure, and membrane curvature is thus more easily estimated.

Assuming a narrow lubrication film exists before the cell fully blocks the channel, we can estimate the optimum velocity by equating the surface tension pressure $2\sigma(1/r_f - 1/R_C)$ to pressure drop in the lubrication film, which includes the pressure drop due to kinetic energy increase, *i.e.* loss in dynamic pressure $\rho V_A^2/2$ and the viscous loss in the liquid film between the cell and the channel wall. $V_A$ is flow velocity at the CTC front. Assuming the thickness of the lubrication film is ε, and the viscous shear stress in the channel can be estimated to be as $\tau = \mu V L/\epsilon$, with an area along the flow direction $A_f = 2\pi R_C L$ and an area perpendicular to the flow direction $A_\odot = 2\pi R_C \varepsilon$. Therefore, the pressure drop caused by the lubrication film is estimated by $\mu Q L^2 / 2\pi R_C \varepsilon^2$.

By combining the relation of the static pressure, lubrication film pressure drop and the dynamic pressure of the carrier flow, we can obtain the following equation for optimum velocity,

$$\frac{\rho(V_A)^2}{2} + \frac{\mu Q L^2}{2\pi R_C \varepsilon^2} \cong 2\sigma(1/r_f - 1/R_{CTC}) \tag{7}$$

At a velocity just high enough to cause CTC yield to carrier fluid, the pressure is the highest critical pressure and flow rate $Q$ in the film is zero. Assuming fully developed flow in the confinement channel, the velocity at the front of CTC $V_A$ along the centreline of the channel is twice the average velocity in the confinement channel.

From Equation (7), we can see the optimum velocity (average velocity $\bar{V}_{cr}$ in the confinement channel) for soft CTC can be described by the following equation,

$$\bar{V}_{cr} = \sqrt{\frac{\sigma}{\rho}\left(\frac{1}{R_C} - \frac{1}{R_{CTC}}\right)} \tag{8}$$

When the velocity is higher than optimum velocity, we predict the existence of co-flow of CTC and carrier fluid and a decreasing of CTC filtration pressure. Theoretical relation in Fig. 7 is plotted from Eq (8).

In Fig. 8, we plotted the phase chart again to indicate typical operational parameter of a deformation based CTC microfilter. The optimum velocity should be a vertical line in the phase chart. Indeed, for the CTC of the parameter of 16μm, microfilter channel of the diameter of 5 *μm*, the critical *We* number can be calculated by Eq (4) and Eq (8), $We = 2 \times (1 - R_C/R_{CTC})$. For our case $We_c$=1.375, which indicates the transition happens when inertial force is comparable to surface tension force. However, our model becomes less accurate for extremely soft cells (white blood cell) when searching for the optimum velocity. This is probably due to too small a value for surface tension and errors involved in the detachment criterion (as explained in Fig 7 caption).

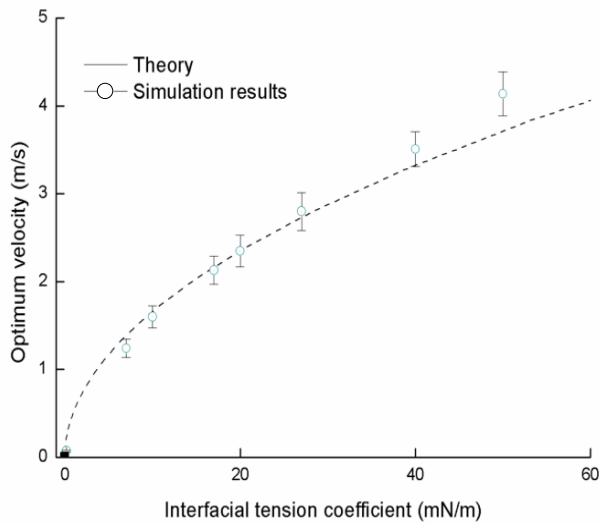

Fig. 7 Surface tension-optimum velocity relation. The velocity is obtained by repeated determination of two velocities. $V^-<V_{cr}<V^+$, $V^-$ represents the velocity at which a cell is squeezed through the channel, $V^+$ represents the velocity at which the a cell is sheared through the channel. The criterion for detachment is selected as lubrication film less than one mesh thickness near the wall of filter.

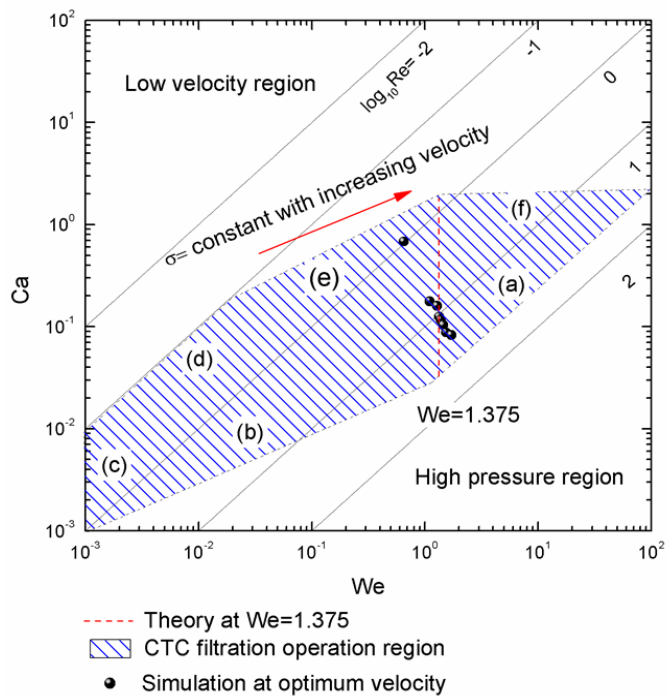

Fig. 8 CTC detection operation chart. Middle hardness and hard cell fits $We=1.375$ well. However the soft CTC has smaller optimum velocity value.

There are several physical and engineering constrains for CTC lab-on-a-chip device design,

1) The total pressure should be smaller than 200kpa for typical microfluidic bonding strength, which limits the operating flow rate;

2) Acceptable time duration (ideally < 1h) for processing a standard sized blood sample (8 mL) [60]

3) Interfacial tension coefficient value of different cells.

These constrains gives the shaded area on the flow regime chart in Fig. 8, where the vertical dashed line indicates optimum flow velocity to obtain the highest maximum passing pressure.

Further derivation yields following expression among dimensionless numbers, which helps define the boundaries of (b) and (e) in Fig. 8,

$$lg(Ca) = \frac{1}{2}lg(We) - \frac{1}{2}lg(\sigma) - \frac{1}{2}lg(\frac{\mu^2}{\rho d}) \tag{9}$$

Equation (10) indicates that contours of constant $\sigma$, i.e., a specific cell type, are tilted lines with a slope of ½, which helps explain the data distribution in Fig. 4.

We can get the corresponding boundary under the confinement of the following $Ca$, $We$, $Re$ limit due to the confinement above. $Re$ spans the range of $Re \in (10^{-2}, 10^2)$, suggesting the time duration of processing time. The lower limit reflects slow processing tolerance and the high $Re$ reflects a high pressure limit by limitation (1).

## D. EFFECTS OF 3D GEOMETRY ON MAXIMUM PASSING PRESSURE

Previous results show that due to the shear flow the Young-Laplace equation does not accurately describe a non-circular channel. Channel geometry was found to be influential on static Laplace pressure prediction for dynamics process [16]. The influence of channel geometry on pressure[31] and deformation[32] in a low speed dynamic process is investigated for circular, square, triangular, and rectangular[16] channel cross-sections. Circular[61], square[51, 62, 63] and rectangular cross-section channel have been extensively investigated but triangular channel have not been well-studied due to difficulties in manufacturing.

Based on the Young-Laplace analysis, the maximum passing pressure for a non-circular channel is given by[59]:

$$P_{sur} = \sigma(\frac{C_{Pw}}{A_P} - \frac{2}{R_{cell}}) \tag{10}$$

$C_{pw}$ is the wet perimeter of cross-section; $A_P$ is the area of cross-section. Theoretical calculation of critical surface tension can be obtained from previous reference.

The values of passing pressure for a cell entering a microchannel of particular channel cross-section are plotted in Fig. 9. The x-axis is the normalized time for ease of comparison.

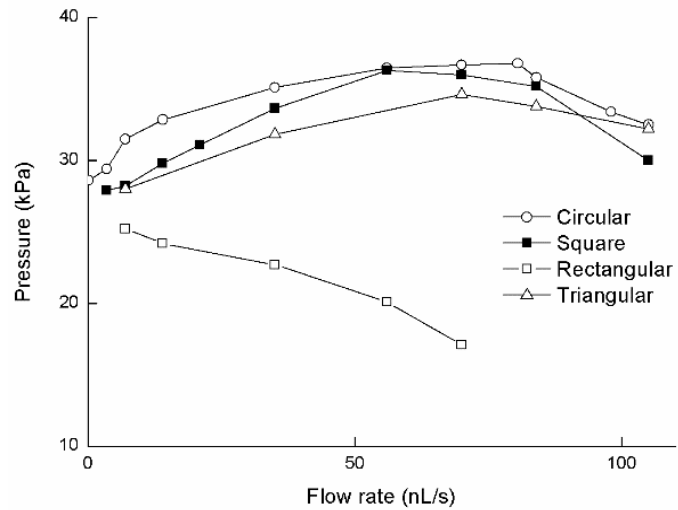

Fig. 9 Optimum velocity-flow rate relation for four microchannel cross-sections: circular, square, triangular and rectangular for a CTC with interfacial tension coefficient of 50 mN/m.

As seen in Fig 9, optimum velocity is observed to influence the filtration pressure of CTC entering a microfilter channels. For these three types of channels, maximum passing pressure increases first with increasing velocity and then decreases after the optimum velocity. However, such behavior is not observed in a rectangular channel with an aspect ratio of 2. For this rectangular cross-section, the pressure continues to decrease with increasing inlet velocity; no optimum velocity is observed asin the other cross-sections.This is due to a lubrication film[64, 65] of carrier fluid between the cell and wall. It is filled with the carrier fluid and confined by the cell and wall. As seen in Fig. 10, due to the anisotropic elongation of the cell body, the cell is more easily compressed from the wider side of the channel. Thus the lubrication film becomes larger and pressure in the channel drops. Consequently, the cell may pass more easily through the microfilter.

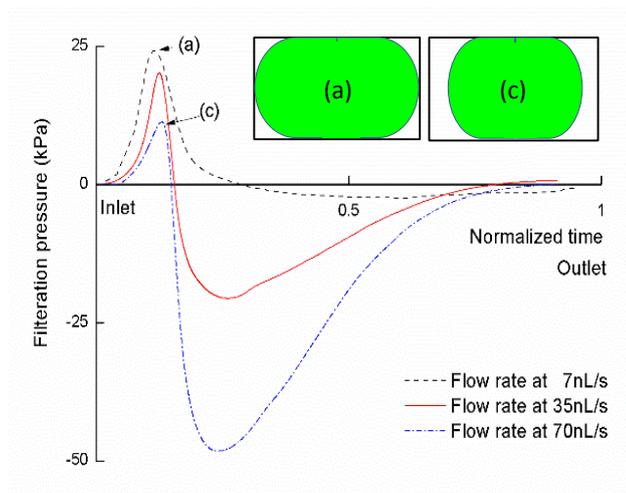

Fig. 10 Dynamic pressure signature for of a CTC passing through a rectangular microchannel with aspect ratio of 2. The x-axis is the normalized time: 0 stands for the starting point of cell entering, 1 stands for the cell recovery to sphere. The y-axis is the filtration pressure, which is the difference between inlet pressure and the viscous pressure drop.

Rather than roundness, microchannel can be characterized by the order of rotational symmetry- the number of equal sides a regular polygon hasFor a regular polygon, channels with an order of rotational symmetry greater than 2 have an advantage in CTC filtering.

## V. Conclusion

In this study, the transition condition of soft matter CTC is conducted numerically using the droplet model. We have demonstrated the deformation evolution of CTC in various microchannel confinements and under different flow conditions. A phase chart for CTC deformation is obtained. The velocity variation line for a specific cell is found to be linear in *We-Ca* log-log coordinates with slope of ½. Also, two different flow regimes of CTC entering a confined microchannel are observed: 1) cell squeezing with no carrier fluid lubrication film and 2) cell shearing with co-flow of carrier fluid. Correspondingly, there are mainly three types of CTC profile at critical pressure 1) hemisphere CTC front with CTC fully blocks channel inlet, 2) parabolic profile at higher speed with also fully blocked inlet, 3) extreme elongation of CTC with a surrounding carrier flow. Moreover, the maximum passing pressure is also observed to vary with the increase of inlet velocity. For circular channel and square channel, the maximum passing pressure increases first then decreases after optimum velocity. However, for a rectangular channel, the maximum passing pressure decreases continuously with the increase of inlet velocity. No optimum velocity is observed. Finally, the optimum velocity is discussed and expressed with numerical verification.

## V. Acknowledgement
We thank Adam Popma for proofreading the manuscript.## VI. Reference

1. R. Siegel, J. Ma, Z. Zou, andA. Jemal, "Cancer statistics, 2014," CA: A cancer journal for clinicians **64**, 9 (2014).
2. J. S. Goodwin, W. C. Hunt, C. R. Key, andJ. M. Samet, "The effect of marital status on stage, treatment, and survival of cancer patients," The Journal of the American Medical Association **258**, 3125 (1987).
3. D. Saslow, C. D. Runowicz, D. Solomon, A. B. Moscicki, R. A. Smith, H. J. Eyre, andC. Cohen, "American Cancer Society guideline for the early detection of cervical neoplasia and cancer," CA: A cancer journal for clinicians **52**, 342 (2002).
4. J.-E. Johansson, O. Andrén, S.-O. Andersson, P. W. Dickman, L. Holmberg, A. Magnuson, andH.-O. Adami, "Natural history of early, localized prostate cancer," The Journal of the American Medical Association **291**, 2713 (2004).
5. S. F. Sener, A. Fremgen, H. R. Menck, andD. P. Winchester, "Pancreatic cancer: a report of treatment and survival trends for 100,313 patients diagnosed from 1985–1995, using the National Cancer Database," Journal of the American College of Surgeons **189**, 1 (1999).
6. F. Fabbri, S. Carloni, W. Zoli, P. Ulivi, G. Gallerani, P. Fici, E. Chiadini, A. Passardi, G. L. Frassineti, andA. Ragazzini, "Detection and recovery of circulating colon cancer cells using a dielectrophoresis-based device: KRAS mutation status in pure CTCs," Cancer Letters **335**, 225 (2013).
7. C. Alix-Panabières, andK. Pantel, "Technologies for detection of circulating tumor cells: facts and vision," Lab on a Chip **14**, 57 (2014).
8. B. Hong, andY. Zu, "Detecting Circulating Tumor Cells: Current Challenges and New Trends," Theranostics **3**, 377 (2013).
9. M.-Y. Kim, T. Oskarsson, S. Acharyya, D. X. Nguyen, X. H.-F. Zhang, L. Norton, andJ. Massague, "Tumor self-seeding by circulating cancer cells," Cell **139**, 1315 (2009).
10. S. C. Hur, N. K. Henderson-MacLennan, E. R. McCabe, andD. Di Carlo, "Deformability-based cell classification and enrichment using inertial microfluidics," Lab on a Chip **11**, 912 (2011).
11. X. Ding, Z. Peng, S.-C. S. Lin, M. Geri, S. Li, P. Li, Y. Chen, M. Dao, S. Suresh, andT. J. Huang, "Cell separation using tilted-angle standing surface acoustic waves," Proceedings of the National Academy of Sciences **111**, 12992 (2014).**15**